\documentclass{ws-p10x7}
\usepackage{times,hyperref}

\begin{document}

\title{CONFINEMENT AND FLAVOR SYMMETRY BREAKING\\VIA MONOPOLE 
CONDENSATION}

% \author{Giuseppe Carlino}
% 
% \address{Department of Physics, University of Wales Swansea,
% Swansea, SA2 8PP, UK\\E-mail: g.carlino@swansea.ac.uk}
% 
% \author{Ken Konishi}
% 
% \address{Dipartimento di Fisica, Universit\`a di Pisa, Sezione di Pisa\\
% E-mail: konishi@mail.df.unipi.it}  
% 
% \address{Istituto Nazionale di Fisica Nucleare, 
% Via Buonarroti, 2, Ed.B, 56127 Pisa, Italy}
% 
\author{HITOSHI MURAYAMA}

\address{Department of Physics, University of California, Berkeley
CA 94720, USA}

\address{Lawrence Berkeley National Laboratory, Cyclotron Road,
Berkeley CA94720 USA\\ E-mail: murayama@lbl.gov}

\twocolumn[\maketitle\abstract{We discuss dynamics of $N=2$ 
supersymmetric $SU(n_{c})$ gauge theories with $n_{f}$ quark 
hypermultiplets.  Upon $N=1$ perturbation of introducing a finite mass 
for the adjoint chiral multiplet, we show that the flavor $U(n_{f})$ 
symmetry is dynamically broken to $U(r) \times U(n_{f}-r)$, where 
$r\leq [n_{f}/2]$ is an integer.  This flavor symmetry breaking occurs 
due to the condensates of magnetic degrees of freedom which acquire 
flavor quantum numbers due to the quark zero modes.  We briefly 
comment on the $USp(2n_{c})$ gauge theories.  This talk is based on 
works with Giuseppe Carlino and Ken Konishi.\cite{CKM1,CKM2}}]

\section{Introduction}%1

There have been at least two main dynamical issues in gauge theories: 
confinement and flavor symmetry breaking.  The former is an obvious 
requirement in understanding the real-world strong interaction 
dynamics, namely the lack of observation of isolated quarks.  The 
spectrum of light hadrons demands a linear potential with respect to 
the distance beween the quark and the anti-quark in meson boundstates.  
The latter is a more subtle issue.  Nambu pointed out that the 
lightness of the pions can be understood if they are what we now call 
Nambu--Godstone bosons of spontaneously broken symmetries.  We need 
the $SU(3)_{L}\times SU(3)_{R}$ flavor symmetry of the QCD to be 
dynamically broken down to $SU(3)_{V}$ by quark bilinear condensates
\begin{equation}
	\langle \bar{u} u \rangle = \langle \bar{d} d \rangle
	= \langle \bar{s} s \rangle \neq 0 .
\end{equation}

An important question is what microscopic mechanism is behind the 
confinement and dynamical flavor symmetry breaking.  The seminal 
work by Seiberg and Witten\cite{SW1} showed that, using $N=2$ 
supersymmetric gauge theories, confinement can be understood as a 
consequence of the magnetic monopole condensation as conjectured a long 
time ago by `t Hooft and Mandelstam.\cite{conjecture} Our aim is to 
bring the understanding of the dynamical flavor symmetry breaking to 
the same level, done in collaboration with Giuseppe Carlino and Ken 
Konishi\cite{CKM1,CKM2}.  Surprisingly, this question had not been 
addressed systematically so far.  Seiberg and Witten themselves 
studied the case with flavor,\cite{SW2} but there were only two 
examples which exhibited dynamical flavor symmetry breaking ($SU(2)$ 
with $n_{f}=2,3$) and it was not possible to draw a general lesson.  
Later works on general gauge groups\cite{APSe,APSh} focused on the 
appearance of the dual gauge group, and did not discuss the issue of 
flavor symmetry breaking.

We start with $N=2$ supersymmetric $SU(n_{c})$ QCD with $n_{f}$ 
hypermultiplet quarks in the fundamental representation.  We later 
add a perturbation which leaves only $N=1$ supersymmetry, $W = \mu 
{\rm tr}\Phi^{2}$, a mass term for the adjoint chiral superfield in 
the $N=2$ vector multiplet.  This theory has $U(n_{f})$ flavor 
symmetry.  We found that the flavor symmetry is in general dynamically 
broken as
\begin{equation}
	U(n_{f}) \rightarrow U(r) \times U(n_{f}-r).
\end{equation}
There are isolated vacua for $0 \leq r \leq [n_{f}/2]$.  We have shown
that this dynamical flavor symmetry breaking is caused by condensation
of magnetic degrees of freedom.  For the vacuum $r=0$, there is no
breaking of the flavor $U(n_{f})$ symmetry.  For the vacuum $r=1$, what
condenses is nothing but the magnetic monopoles, which belong to the
fundamental representatin of the $U(n_{f})$ flavor group.  For the
vacua $r>1$, magnetic monopoles ``break up'' into ``dual quarks''
before reaching the singularities where they become massless; it is
the ``dual quark'' which condenses and breaks the flavor symmetry.  In
any case, the flavor symmetry breaking and the confinement\footnote{We 
use the term ``confinement'' somewhat loosely, as in ``s-confinement'' 
in \cite{s-confinement}.} are both
caused by the condensation of magnetic degerees of freedom.

Thanks to holomorphy, there is no phase transition by varying $\mu$ 
from small ($\mu \ll \Lambda$) to large ($\mu \gg \Lambda$).  
Therefore one can study the theory in both limits and compare the 
results; this would not only provide us non-trivial cross checks but 
also insight into the dynamics of the theory.  We can also resort to 
completely different techniques to analyze the theory in the different 
limits.

In the limit (1) ($\mu \gg \Lambda$), we can integrate the adjoint chiral 
multiplet $\Phi$ out from the theory, and study the resulting $N=1$ 
low-energy theory.  The low-energy theory has a superpotential term
\begin{equation}
	W = - \frac{1}{\mu} (\tilde{Q} T^{a} Q) (\tilde{Q} T^{a} Q),
	\label{eq:Weff}
\end{equation}
where $Q$ ($\tilde{Q}$) are the quark chiral superfields in the 
fundamental (anti-fundamental) representation of the gauge group.  
Then we can use analysis by Seiberg\cite{Seiberg} on $N=1$ 
supersymmetric QCD together with the above effective superpotential 
(\ref{eq:Weff}).  It is then easy to identify the vacua of the theory 
by solving for the extrema of the superpotential with respect to the 
gauge-invariant composites such as $M^{ij} = \tilde{Q}^{i} Q^{j}$ or 
$B^{i_{1} \cdots i_{n_{c}}} = Q^{i_{1}} Q^{i_{2}} \cdots 
Q^{i_{n_{c}}}$, $\tilde{B}^{i_{1} \cdots i_{n_{c}}} = 
\tilde{Q}^{i_{1}} \tilde{Q}^{i_{2}} \cdots \tilde{Q}^{i_{n_{c}}}$.  
This makes it easy to identify the flavor symmetry breaking patterns. 

In the other limit (2) ($\mu \ll \Lambda$), we start with $N=2$ limit 
($\mu=0$) where the low-energy effective theory is known exactly.  In 
this limit, we can identify monopole degrees of freedom etc which 
become massless at singularities.  We then turn on $\mu \neq 0$.  This 
way we obtain information on the microscopic dynamics of magnetic 
degrees of freedom.

When considering the theory in various limits, a very powerful check
is provided by counting the number of vacua, by further perturbing the
theory by finite masses of hypermultiplet quarks.  Quark masses make
the vacua discrete and countable, and we must obtain the same number
of vacua in different limits.  In fact, we considered four such limits
in total.  Two of them have large $\mu$.  In the limit (1A), we regard
both $\mu$ and $m_{i}$ large and solve for vacua semi-classically
({\it i.e.}\/, including the effects of gaugino condensates in
unbroken pure Yang-Mills factors).  In the limit (1B), we integrate
out $\Phi$ and use known $N=1$ dynamics together with the effective
superpotential Eq.~(\ref{eq:Weff}), further combined with the mass
terms for the quarks.  The other two have small $\mu$, namely setting
$\mu=0$ first, and then reintroduce $\mu \neq 0$ later on.  In the
limit (2A), we approach singularities from large $\Phi$ on the Coulomb
branch.  In the limit (2B), we approach singularities from large $Q$,
$\tilde{Q}$ on the Higgs branch.  All these approaches should give the
identical number of vacua, and the consistency among them tell us, for
example, which singularity on the Coulomb branch corresponds to which
symmetry breaking pattern.

I will not discuss the limit (1A) in this talk and simply refer
interested parties to our paper.\cite{CKM2} I first discuss the limit
(1B) and identify the flavor symmetry breaking patterns.  Then I will
briefly review how the monopoles acquire flavor quantum numbers.  I
will move on to the analyses with small $\mu$ next, first on the
Coulomb branch, and next on the Higgs branch.  The consistency among
different approaches gives us full understanding of the dynamics.

\section{Large $\mu$ Analysis}

$N=2$ supersymmetric QCD can be viewed as a special version of $N=1$ 
supersymmetric gauge theories with the following superpotential
\begin{equation}
	W = \sqrt{2} \tilde{Q}_{i} \Phi Q_{i} + m_{i} \tilde{Q}_{i} Q_{i}
		+ \mu {\rm tr}\Phi^{2},
\end{equation}
where the last term breaks $N=2$ to $N=1$.  When $\mu$ is large, we 
can integrate out $\Phi$ field, and obtain
\begin{equation}
	W = - \frac{1}{\mu} (\tilde{Q}_{i} T^{a} Q_{i}) (\tilde{Q}_{j} T^{a} Q_{j})
		+ m_{i} \tilde{Q}_{i} Q_{i}.
\end{equation}
Doing Fierz transformation on the first term, we obtain
\begin{equation}
	W = \frac{1}{2\mu} \left[ {\rm tr}M^{2} - \frac{1}{n_{c}}({\rm tr}M)^{2}
		\right] + {\rm tr} m M,
		\label{eq:WM}
\end{equation}
where $M_{ij} = \tilde{Q}_{i} Q_{j}$ is the meson chiral superfield,
and the mass $m = {\rm diag}(m_{1},\cdots, m_{n_{f}})$ and the meson 
field are in the matrix notation. 

Due to lack of time, I concentrate on the case $n_{f} < n_{c}$.  I
refer to our papers\cite{CKM1,CKM2} for larger number of flavors.  In
this case, the non-perturbative superpotential\cite{ADS} is added to
Eq.~(\ref{eq:WM}):
\begin{equation}
	\Delta W = (n_{c}-n_{f}) \frac{\Lambda_{1}^{(3n_{c}-n_{f})/(n_{c}-n_{f})}}
		{({\rm det}M)^{1/(n_{c}-n_{f})}},
	\label{eq:WADS}
\end{equation}
where $\Lambda_{1}^{3n_{c}-n_{f}}=\mu^{n_{c}}\Lambda^{2n_{c}-n_{f}}$ 
is the scale of the low-energy $N=1$ theory.

By solving for the meson matrix $M = {\rm diag}(\lambda_{1}, \cdots, 
\lambda_{n_{f}})$, we find that $\lambda_{i}$ satisfy quadratic 
equations and hence there are two solutions for each of them.  We 
obtain
\begin{equation}
	\lambda_{i} = \frac{1}{2} (Y \pm \sqrt{Y^{2} + 4\mu X}) + O(m),
\end{equation}
where the signs $\pm$ indicate two solutions for each $i = 1, \cdots, 
n_{f}$ and hence there are $2^{n_{f}}$ possibilities.  For the choice 
of $r$ plus signs and $n_{f}-r$ minus signs, we can further determine 
$X$ and $Y$, which can take $(2n_{c}-n_{f})$ possible phases.  
Avoiding double counting for $r \leftrightarrow n_{f}-r$, we find 
$(2n_{c}-n_{f})2^{n_{f}-1}$ vacua in total.  The most important 
outcome from this analysis is that, in $m\rightarrow 0$ limit, $r$ 
eigenvalues with plus sign are degenerate, and $n_{f}-r$ eigenvalues 
with minus sign are also degenerate.   Such a vacuum for the meson 
field exhibits dynamical flavor symmetry breaking, $U(n_{f}) 
\rightarrow U(r) \times U(n_{f}-r)$.  

\section{Semi-classical Monopoles}

As was shown by `t Hooft and Polyakov, there are solitonic solutions 
to the gauge-Higgs system which appear as magnetic monopoles under the 
low-energy gauge group.  The canonical example is the $SU(2)$ gauge 
theory with the adjoint Higgs $\Phi$, where the expectation value of 
$\Phi = a \sigma_{3}$ breaks $SU(2) \rightarrow U(1)$.  The mass of 
the magnetic monopole is given roughly as $M \sim 4\pi a/g$, while the 
mass of the $W$-boson is $m_{W} \sim g a$.  Therefore for the 
weakly-coupled case, the magnetic monopole is heavy and $W$ (electric 
monopole) is light, while for the strongly-coupled case, the magnetic 
monopole is light and the $W$-boson is heavy.

The case with flavor is quite interesting.\cite{flavor} The
cancellation of the $SU(2)$ Witten anomaly requires an even number of
flavors: $2n_{f}$ doublet quarks.  They can couple to the adjoint
Higgs as ${\cal L}_{\rm Yukawa} = q_{i} \Phi q_{i}$, which produces
Majorana-type mass terms.  The largest possible flavor symmetry is
$SO(2n_{f})$.  Solving Dirac equation for the quarks in the monopole
background, there is one zero-energy mode for each flavor.  The
important question is what statistics the fermion zero modes follow. 
Surprisingly, they are bosons.  The reasoning is simple.  The way to
judge if an excitation is bosonic or fermionic is by studying the
$2\pi$ rotation of space and asking if the excitation produces a minus
sign (fermion) or not (boson).  In the presence of a `t
Hooft--Polyakov monopole, a naive $2\pi$ spatial rotation is not a
symmetry because the isospin space is tied to the real space
(``hedgehog'').  Therefore one needs to make a $2\pi$ rotation both
for the real space and the isospin (gauge) space to determine
statistics.  For fermion zero modes in the $SU(2)$ doublet
representation, spatial rotation produces a minus sign, while the
isospin rotation produces another minus sign.  The fermion zero mode
does not produce a minus sign under the true $2\pi$ rotation and hence
is a boson.  Therefore monopole states with or without the fermion
zero mode have the same statistics and the same energy.  In other
words, the monopole states form a multiplet.

For the $SU(2)$ gauge theory, or in general $USp(2n_{c})$ gauge 
theories, we have $SO(2n_{f})$ flavor symmetry.  Fermion zero mode 
operators $q^{i}$ follow the anti-commutation relation 
$\{q^{i},q^{j}\} = \delta^{ij}$ upon canonical quantization, and they 
are represented as gamma matrices $q^{i} = \gamma^{i}/\sqrt{2}$.  The 
monopole Hilbert space is the representation space of the 
anti-commutation relation, and is hence a spinor representation of 
$SO(2n_{f})$, with $2^{n_{f}}$ states.  For the $SU(n_{c})$ gauge 
theories ($n_{c} > 2$), however, the flavor symmetry is only as large 
as $U(n_{f}) \subset SO(2n_{f})$, and hence the monopole multiplet 
(spinor under $SO(2n_{f})$) is decomposed into irreducible multiplets 
under $U(n_{f})$: totally anti-symmetric tensor representations.  One 
can easily check that the the dimensions match: $\sum_{r} {}_{n_{f}} 
C_{r} = 2^{n_{f}}$ using the binomial theorem.

We have learned that the monopoles acquire flavor quantum numbers of
the rank-$r$ totally anti-symmetric representations, while the theory
breaks the $U(n_{f})$ flavor symmetry dynamically to $U(r) \times
U(n_{f}-r)$ in the previous section.  We are naturally led to a
conjecture that the flavor symmetry breaking is caused by the
condensation of the magnetic monopoles which causes confinement at the
same time.

\section{Moduli Space of $N=2$ Theories}

The classical moduli space of the theory is determined by solving the 
vacuum equations
\begin{eqnarray}
	\Phi Q_{i} & = & 0 ,
	\label{eq:Qtilde}  \\
	\tilde{Q}_{i} \Phi & = & 0 ,
	\label{eq:Q}  \\
	\sum_{i} \left\{ Q_{i} \tilde{Q}_{i} 
		- \frac{1}{n_{c}} {\rm tr}Q_{i} \tilde{Q}_{i} \right\} & = & 0 ,
	\label{eq:Phi}  \\
	\left[\Phi, \Phi^{\dagger}\right] & = & 0 ,
	\label{eq:D1}  \\
	Q^{\dagger} T^{a} Q - \tilde{Q}^{T} T^{a} \tilde{Q}^{*} & = & 0 .
	\label{eq:D2}
\end{eqnarray}

There are three types of ``branches'' to the vacuum 
solutions\cite{APSe}: (1) Coulomb branch, (2) Non-baryonic (or mixed) 
branch, and (3) Baryonic branch.  The baryonic branch appears only for 
$n_{f} \geq n_{c}$ and we will not discuss it.

The solution to Eq.~(\ref{eq:D1}) is given by $\Phi = {\rm 
diag}(\phi_{1}, \cdots, \phi_{n_{c}})$ with the constraint ${\rm 
tr}\Phi = \sum_{k} \phi_{k} = 0$.  This defines the complex 
$(n_{c}-1)$-dimensional Coulomb branch.  At a generic point on the Coulomb 
branch, the theory is a free $U(1)^{n_{c}-1}$ gauge theory while 
there appear massless particles on singular submanifolds.  The 
singularities can be found where the auxiliary curve\cite{curve}
\begin{equation}
	y^{2} = \prod_{k=1}^{n_{c}}(x-\phi_{k})^{2}
		+ 4 \Lambda^{2n_{c}-n_{f}}\prod_{i=1}^{n_{f}}(x+m_{i})
	\label{eq:curve}
\end{equation}
is maximally degenerate.

The non-baryonic branch is given by the following field 
configurations:
\begin{eqnarray}
	Q & = & \left(
		\begin{array}{ccc|ccc|c}
			\kappa_{1} & & & 0 & & & 0\\
			& \ddots & & & \ddots & & \vdots \\
			& & \kappa_{r} & & & 0 & 0\\ \hline
			0 & & & 0 & & & 0\\
			& \ddots & & & \ddots & & \vdots \\
			& & 0 & & & 0 & 0
		\end{array}
		\right)
	\label{eq:Q0}  \\
	\tilde{Q} & = & \left(
		\begin{array}{ccc|ccc|c}
			0 & & & \kappa_{1} & & & 0\\
			& \ddots & & & \ddots & & \vdots \\
			& & 0 & & & \kappa_{r} & 0\\ \hline
			0 & & & 0 & & & 0\\
			& \ddots & & & \ddots & & \vdots \\
			& & 0 & & & 0 & 0
		\end{array}
		\right)
	\label{eq:Qtilde0}  \\
	\Phi & = & \left( 
		\begin{array}{ccc|ccc}
			0 & & & & & \\
			& \ddots & & & 0 & \\
			& & 0 & & & \\ \hline
			& & & \phi_{r+1} & & \\
			& 0 & & & \ddots & \\
			& & & & & \phi_{n_{c}}
		\end{array}
		\right)
	\label{eq:Phi0}
\end{eqnarray}
Because both hypermultiplets $Q$, $\tilde{Q}$ and the vector multiplet
$\Phi$ have expectation values, it is also called the mixed branch. 
There are separate $r$-branches for each choice of the integer $r$. 
The integer $r$ can range from 1 to ${\rm min}\{ [\frac{n_{f}}{2}],
n_{c}-2\}$, and hence this branch exists only for $n_{f}\geq 2$ and
$n_{c}\geq 3$.  It is important to note that the limit $\kappa_{k}
\rightarrow 0$ recovers $U(r)$ gauge symmetry and the branch touches
the Coulomb branch (``root'' of the no-baronic branch).  The theory at
the root is a $U(r) \times U(1)^{n_{c}-r-1}$ gauge theory which is
asymptotically non-free, and hence the gauge fields survive as
dynamical degrees of freedom in the low-energy limit.  Along the root,
there are special isolated points where we can find $n_{c}-r-1$
massless monopole multiplets so that the curve becomes maximally
degenerate.

\section{Coulomb Branch Description}

One can locate points on the Coulomb branch where the curve is 
maximally degenerate, so that the points survive after $\mu \neq 0$ 
perturbation.  They do lie on the roots of $r$-branches.  After mass 
perturbation for the hypermultiplets, one can count the number of 
vacua by identifing the points on the Coulomb brach which coalese to 
the same point when the masses are turned off.  This is a technically 
involved analysis which required us many pages of the paper 
\cite{CKM2}.  Nonetheless the result is simple.  Starting from the 
maximally degenerate point on the $r$-branch root, the mass 
perturbation splits the point into ${}_{n_{f}}C_{r}$ vacua.  Comparing 
this counting to the large $\mu$ analysis, we can say that the vacuum 
at the $r$-branch root breaks the flavor symmetry as $U(n_{f}) 
\rightarrow U(r) \times U(n_{f}-r)$.

Therefore, the following picture appears true.  The semi-classical 
monopoles far away from the singularities on the Coulomb branch 
acquire the flavor quantum number of the rank-$r$ totally 
anti-symmetric tensor prepresentation under the $U(n_{f})$ flavor 
group. They become massless at the maximally degenerate point along the 
$r$-branch root and condense upon $\mu \neq 0$ perturbation.  This 
picture, however, leaves a paradoxical situation.  The low-energy 
effective Lagrangian of the monopoles would have a large accidental 
symmetry $U(_{n_{f}}C_{r})$, and upon condensation of one of the 
components, all the others remain massless.  Even though it is 
logically not impossible, it casts some doubts about this naive 
picture.  Indeed, something non-trivial happens between the 
semi-classical regime and the $r$-branch root as I will describe in 
the next section.

\section{Low-energy Effective Lagrangians}

Now we aproach the singularities from the Higgs branch.  As remarked 
earlier, by turning off $\kappa_{k} \rightarrow 0$, we can approach 
the roots where we recover infrared-free $U(r)\times U(1)^{n_{c}-r-1}$ 
gauge theory.  At the maximally degenerate point along the roots, we 
have also $n_{c}-r-1$ additional magnetic monopole hypermultiplets 
$e_{k}$, $\tilde{e}_{k}$ coupled to the $U(1)$ factors.  The 
fundamental quarks still couple to the $SU(r)$ gauge factor as the 
fundamental representation $q_{i}$ and $\tilde{q}_{i}$ because the 
non-renormalization theorem guarantees the flat hyper-K\"ahler metric 
for the quarks.  The unique effective Lagrangian obtained this way 
is\cite{APSe}
\begin{equation}
	W = \sqrt{2} \tilde{q}_{i}\phi q_{i} 
	+ \sqrt{2} \psi_{0} \tilde{q}_{i} q_{i}
	+ \sqrt{2} \sum_{k=1}^{n_{c}-r-1} \psi_{k} \tilde{e}_{k} e_{k},
\end{equation}
where $\phi$, $\psi_{0}$ belong to the $U(r)$ vector multiplet and 
$\psi_{k}$ to each of the $U(1)$ vector multiplets.  The $N=1$ 
perturbation then is given by
\begin{equation}
	\Delta W = \mu \Lambda \sum_{j=0}^{n_{c}-r-1} x_{j} \psi_{j}
		+ \mu {\rm tr}\phi^{2},
\end{equation}
where $x_{j}$ are $O(1)$ constants, and we find vacua
\begin{eqnarray}
	 &  & q=\tilde{q}=\left( \begin{array}{ccc|c}
	 	1 & & & 0\\
		& \ddots & & \vdots\\
		& & 1 & 0
	 \end{array}
	 \right)\sqrt{-\frac{\mu\Lambda}{\sqrt{2}r}},
	  \\
	 &  & e_{k}=\tilde{e}_{k}=\sqrt{-\mu\Lambda},
	 \quad \psi_{0}=\psi_{k}=0.
\end{eqnarray}
Note that the expectation values of $q$, $\tilde{q}$ break the flavor 
symmetry $U(n_{f})$ to $U(r) \times U(n_{f}-r)$, where the unbroken 
$U(r)$ is the diagonal subgroup of the flavor group $U(n_{f})$ and the 
$U(r)$ gauge group.  And there are ${}_{n_{f}}C_{r}$ choices to pick 
$r$ flavors out of $n_{f}$ quark flavors for vacuum expectation values.

The semi-classical monopoles in the rank-$r$ anti-symmetric tensor 
representation therefore must have ``broken up'' into ``dual quarks'' 
in the way that the monopole $M_{i_{1}\cdots i_{r}} = q_{i_{1}} \cdots 
q_{i_{r}}$ is matched to the baryonic composite before one reaches the 
singularities on the Coulomb branch.  For the special case of $r=1$, 
the ``dual quarks'' themselves are the magnetic monopoles.  The 
evidence for this identification is the following.  The singularities, 
after $m_{i}\neq 0$ perturbation, have $U(1)^{n_{c}-1}$ gauge 
group and one can study the monodromy around the singularities.  It can 
be seen that there are one massless magnetic monopole for each $U(1)$ 
factors.  By sending quark mass to zero, $_{n_{f}}C_{r}$ singularities 
coalesce into a point where the massless monopoles belong to the 
$U(r)$ quark multiplet.  Therefore, the quarks, which are continuously 
connected to the ``electric quarks'' at the large VEVs along the Higgs 
branches, are indeed magnetic degrees of freedom at the non-baryonic 
branch roots.

\section{Conclusion}

We have studied the issues of confinement and the dynamical flavor 
symmetry breaking in gauge theories, by starting with $N=2$ 
$SU(n_{c})$ QCD with $n_{f}$ flavors and perturbing it by the adjoint 
mass term.  We have shown that both confinement and flavor symmetry 
breaking are caused by a single mechanism: condensation of magnetic 
degrees of freedom which carry flavor quantum numbers.  

We have also studied $USp(2n_{c})$ theories.  There the magnetic 
monopoles are spinors under the $SO(2n_{f})$ flavor group, and cannot 
``break up'' into quarks.  Therefore quarks and monopoles coexist at 
the singularity on the moduli space and the theory becomes 
superconformal.  No local effective Lagrangian can be written and one 
cannot discuss it along the same line as in the $SU(n_{f})$ gauge 
theories.  However, the flavor symmetry is broken to $U(n_{f})$, and 
this is consistent with the condensation of the spinor monopoles.  
This strongly suggests that the overall picture of flavor symmetry 
breaking via monopole condensation is correct in this case as well.

\section*{Acknowledgments}
I thank Beppe Carlino and Ken Konishi for collaboration.  This work 
was supported in part by the Department of Energy under contract 
DE--AC03--76SF00098, and in part by the National Science Foundation 
under grant PHY-95-14797.

\end{document}